\documentclass[11pt,a4paper]{article}
\usepackage{jcappub}
\usepackage[top=1in, bottom=1in, left=1in, right=1in]{geometry}
\usepackage{amsmath, amssymb, amsfonts, mathrsfs, graphicx, fancyhdr,}
\usepackage[singlelinecheck=false]{caption}
\usepackage{subfigure}
\usepackage{setspace}
\usepackage{color}
\usepackage{epstopdf}
\definecolor{darkgreen}{rgb}{0,0.5,0}
\definecolor{darkblue}{rgb}{0,0,0.5}
\usepackage[usenames,dvipsnames]{xcolor}
\usepackage{bbm}
\usepackage{booktabs}
\newcommand{\be}{\begin{equation}}
\def\ee{\end{equation}}

\def\bea{\begin{eqnarray}}
\def\eea{\end{eqnarray}}

\title{Accidental Inflation from K\"ahler Uplifting}

\begin{abstract}
{We analyze the possibility of realizing inflation with a subsequent dS vacuum in the K\"ahler uplifting scenario. The inclusion of several quantum corrections to the 4d effective action evades previous no-go theorems and allows for construction of simple and successful models of string inflation. The predictions of several benchmark models are in accord with current observations, i.e., a red spectral index, negligible non-gaussianity, and spectral distortions similar to the simplest models of inflation. A particularly interesting subclass of models are ``left-rolling" ones, where the overall volume of the compactified dimensions shrinks during inflation. We call this phenomenon ``inflation by deflation" (IBD), where deflation refers to the internal manifold. This subclass has the appealing features of being insensitive to initial conditions, avoiding the overshooting problem, and allowing for observable running $\alpha \sim 0.012$ and enhanced tensor-to-scalar ratio $r \sim 10^{-5}$. The latter results differ significantly from many string inflation models.}  
\end{abstract}

\author{Ido Ben-Dayan$^1$, Shenglin Jing$^{2,3}$, Alexander Westphal$^1$, Clemens Wieck$^1$}
\affiliation{$^1$Deutsches Elektronen-Synchrotron DESY, Theory Group, D-22603 Hamburg, Germany\\
$^2$Canadian Institute for Theoretical Astrophysics, University of Toronto, 60 St.George Street, Toronto, ON, M5S 3H8, Canada\\
$^3$Department of Astronomy and Astrophysics, University of Toronto, 50 St.George Street, Toronto, ON, M5S 3H4, Canada}

\begin{document}
\hfill {DESY-13-155}

\maketitle

%
%%
%%%
%%%%
%%%%%%%%%%%%%%%%%%%%%%%%%%%%%%%%%%%%%%%%%%%%%%%
%%%%%%%%%%%%%%%%%%%%%%%%%%%%%%%%%%%%%%%%%%%%%%%
%%%%%%%%%%%%%%				 %%%%%%%%%%%%%%%%%%%%%%%
%%%%%%%%%%%%%	%         section 1        %%%%%%%%%%%%%%%%%%%%%%%
%%%%%%%%%%%%%%				  %%%%%%%%%%%%%%%%%%%%%%
%%%%%%%%%%%%%%%%%%%%%%%%%%%%%%%%%%%%%%%%%%%%%%%
%%%%%%%%%%%%%%%%%%%%%%%%%%%%%%%%%%%%%%%%%%%%%%%
\section{Introduction}

Cosmology of the very early Universe has seen a spectacular transition from being a largely speculative endeavor before the mid-90s into becoming the subject of high-reliability and increasingly even precision observations largely based on systematic scans for type IA supernovae (SN) and the Cosmic Microwave Background (CMB) radiation. The Hubble space telescope finally yielded reliable estimates for the present-day Hubble parameter $H_0=73.8\pm 2.4 \, \text{km s}^{-1}\text{Mpc}^{-1}$~\cite{Freedman:2000cf,Riess:2011yx}. Next, systematic type IA SN observations led to the detection of a form of dark energy compatible with an extremely tiny ($\sim 10^{-122}M_{\rm P}^4$) yet positive cosmological constant, which drives late-time accelerated expansion of the Universe~\cite{Riess:1998cb,Perlmutter:1998np}. Finally, precision observations of the CMB by the space-based satellite probes WMAP~\cite{Hinshaw:2012fq} and PLANCK~\cite{Ade:2013zuv,Ade:2013uln} as well as the ground-based telescopes ACT~\cite{Sievers:2013wk} and SPT~\cite{Story:2012wx,Hou:2012xq} led to a concordance model of observational cosmology. The `essence' of this host of observations is consistent with certain simple features of our observed Universe. We live to a good approximation (sub-\%-level) in a spatially flat FLRW universe undergoing accelerated late-time expansion driven by dark energy. The CMB provides increasingly precise hints from its power-spectrum of $10^{-5}$-level thermal fluctuations that the very early Universe was subject to an earlier and extremely rapid phase of accelerated expansion driven by the vacuum energy of a slowly-rolling scalar field, called inflation (see e.g.~\cite{Baumann:2009ds} for a recent review).

This progress of the last 15 years forced candidate theories for a fundamental unification of quantum mechanics with Einstein gravitation, such as string theory, to accommodate both an extremely tiny positive cosmological constant and the dynamics of slow-roll inflation for the very early Universe. In particular, string theory solutions are typically subject to a requirement of 10 dimensions of space-time. The six extra dimensions hence need compactification on manifolds with sub-nuclear radii. This leads to an extremely large number of possible string theory vacua due to the very large number of compact manifolds and their deformation modes which appear as 4d scalar `moduli' fields. Crucial progress towards describing late-time and early-time (quasi) de Sitter (dS) stages of positive vacuum energy in string theory involved the construction of string vacua with complete stabilization of all geometrical moduli fields and controllable supersymmetry breaking, largely in type IIB string theory (see e.g.~\cite{Grana:2005jc,Douglas:2006es} for a review), and more recently also in heterotic string theory~\cite{Cicoli:2013rwa}. This led to the appearance of an exponentially large landscape of isolated dS vacua in string theory~\cite{Susskind:2003kw}.

Within these constructions of string theory dS vacua inflation must then arise from a flat region of the moduli scalar potential supporting the criteria of slow-roll. Recent years have seen a large body of work realizing models of inflation in type IIA/IIB string theory which utilize either D-brane positions, volume or K\"ahler moduli, or axions arising from higher-dimensional gauge-fields in string theory as inflaton scalar fields (see e.g.~\cite{Baumann:2009ni,Cicoli:2011zz,Burgess:2013sla} for very recent reviews). 

Most of the type IIB string inflation models constructed so far invoked mechanisms of moduli stabilization involving either the KKLT mechanism~\cite{Kachru:2003aw} or the Large Volume Scenario (LVS)~\cite{Balasubramanian:2005zx,Conlon:2005ki}. All mechanisms invoke three-form fluxes to stabilize the complex structure moduli and type IIB axio-dilaton. KKLT then uses non-perturbative effects to stabilize the volume moduli in a supersymmetric AdS minimum. An additional supersymmetry breaking sector (e.g. an anti-D3-brane, or a matter sector providing F-, or F- and D-terms) provides an uplift to dS. The Large Volume Scenario uses an interplay between non-perturbative effects and the leading ${\cal O}(\alpha'^3)$ correction to the K\"ahler potential of the volume moduli~\cite{Becker:2002nn} to generate a supersymmetry breaking AdS minimum for these moduli at exponentially large volume. Inclusion of D-terms can change these LVS vacua directly to dS~\cite{Cicoli:2012vw,Cicoli:2013mpa}. There is another mechanism called ``K\"ahler uplifting" capable of generating controlled and reasonably explicit dS vacua for type IIB $\text{Calabi--Yau}$ orientifold flux compactifications~\cite{Rummel:2011cd,deAlwis:2011dp,Louis:2012nb}. Like LVS, K\"ahler uplifting uses an interplay between non-perturbative effects and the leading ${\cal O}(\alpha'^3)$ correction to the K\"ahler potential of the volume moduli. However, the F-term of the overall volume modulus provided by the ${\cal O}(\alpha'^3)$ correction is sufficient to directly generate dS vacua.

In this paper we construct models of volume modulus inflation in single-K\"ahler modulus examples of K\"ahler uplifted dS vacua. We find that a single non-perturbative effect cannot support both successful inflation and a graceful exit, while it is sufficient for dS volume stabilization. Consequently, we extend our discussion to examples involving a racetrack of two non-perturbative effects. Using this extension we succeed in uncovering several regimes where the scalar potential of the volume modulus develops a local slow-roll flat saddle point or inflection point driving slow-roll inflation with a graceful exit into the nearby dS minimum of the volume modulus.

All our examples rely on the ability to tune the type IIB Gukov--Vafa--Witten type flux superpotential and the complex structure dependent instanton prefactors for providing a narrow sub-Planckian field range where the slow-roll conditions are satisfied. Consequently, the classical phase space of initial conditions for the volume modulus conducive to entering slow-roll inflation in this narrow field range is parametrically small compared to models of monomial large-field inflation. Hence, our models realize the general notion of `accidental inflation'~\cite{Linde:2007jn} in the concrete setting of K\"ahler uplifted dS vacua.

The discussion proceeds as follows. In section~\ref{sec:setup} we recall the general structure of the effective action for the moduli from type IIB Calabi--Yau orientifold flux compactifications, and the role $\alpha'$-corrections to the volume moduli K\"ahler potential are playing in satisfying the dS/inflation condition on the sectional curvature of the K\"ahler potential~\cite{Covi:2008ea, Covi:2008cn}. We then briefly review the possible CMB observables which may be used to discriminate between inflationary models using e.g. the PLANCK data.

In section~\ref{sec:models} we start with a general argument of continuity, stating that the existence of K\"ahler uplifted dS vacua does imply the existence of slow-roll inflationary saddle or inflection points. However, the second non-perturbative effect of a racetrack setup is necessary to provide the simultaneous co-existence of slow-roll inflation and a dS minimum for a graceful exit from inflation. Analyzing the combination of K\"ahler uplifting with a racetrack superpotential for the volume modulus, we find three regimes of accidental modulus inflation:
\begin{itemize}
\item Inflation is driven by the volume modulus on a local saddle/inflection point. During slow-roll the volume slowly increases (the CY `inflates').
\item Inflation is driven by the volume modulus on a local saddle/inflection point. During slow-roll the volume slowly decreases (the CY `deflates')\footnote{See e.g. the model of fibre inflation~\cite{Cicoli:2008gp} for one of the very few other models where the CY volume `deflates' during inflation.}. By tuning we can arrange this situation to generate an approximate discrete symmetry providing for hill-top model of inflation described to leading order by an inverted quartic potential $V=V_0(1-\lambda \phi_T^4)$ with a field range $\Delta\phi_T\sim 0.3 M_{\rm P}$ for the canonically normalized inflaton field somewhat larger than for typical accidental inflation setups. This in turn allows us to enhance the tensor-to-scalar ratio to levels of $r\sim 10^{-5}$. The achievable enhancement of $r$ is bounded by the amount of running $\alpha\lesssim 0.01$ of the spectral index $n_s$ allowed by data (see e.g.~\cite{Hebecker:2013zda,BenDayan:2009kv,Hotchkiss:2011gz} where similarly the $r$-enhancement is bounded by the allowed running $\alpha$.)
\item Inflation is driven by a combination of the volume modulus and its $C_4$ RR-form axion partner on one of two local symmetrically arranged saddle/inflection points.
\end{itemize}

In all examples the orthogonal direction to the inflationary trajectory is parametrically heavy, so they are effectively single-field models and exhibit negligible non-gaussianity and isocurvature perturbations.
We finally include the effects of a recently derived ${\cal O}(\alpha'^2)$-correction to the volume moduli K\"ahler potential~\cite{Grimm:2013gma} into our analysis. Consistent with~\cite{Pedro:2013qga} we find that our models can tolerate the presence of this new correction if either the coefficient of the correction appears somewhat suppressed (we need $k\lesssim 1$ instead of generically $k={\cal O}(10)$), or if we used significantly larger gauge groups for the racetrack superpotential for moving both the inflationary region and the dS minimum to larger volumes.

Finally, we conclude in section~\ref{sec:concl}, while we relegate some technical details and a discussion of the recently found K\"ahler uplifted racetrack dS minimum of~\cite{Sumitomo:2013vla} to the appendices~\ref{sec:Appendixeom}, and~\ref{sec:Appendixmin}, respectively.

%
%%
%%%
%%%%
%%%%%%%%%%%%%%%%%%%%%%%%%%%%%%%%%%%%%%%%%%%%%%%
%%%%%%%%%%%%%%%%%%%%%%%%%%%%%%%%%%%%%%%%%%%%%%%
%%%%%%%%%%%%%%				 %%%%%%%%%%%%%%%%%%%%%%%
%%%%%%%%%%%%%	%         section 2        %%%%%%%%%%%%%%%%%%%%%%%
%%%%%%%%%%%%%%				  %%%%%%%%%%%%%%%%%%%%%%
%%%%%%%%%%%%%%%%%%%%%%%%%%%%%%%%%%%%%%%%%%%%%%%
%%%%%%%%%%%%%%%%%%%%%%%%%%%%%%%%%%%%%%%%%%%%%%%
\section{K\"ahler moduli stabilization and inflation}\label{sec:setup}
%
%%
%%%
%%%%
%%%%%%%%%%%%%%%%%%%%%%%%%%%%%%%%%%%%%%%%%%%%%%%
%%%%%%%%%%%%%%%%%%%%%%%%%%%%%%%%%%%%%%%%%%%%%%%
\subsection{4d effective action from type IIB flux compactifications}
In generic compactifications of type IIB string theory with three-form fluxes and D7-branes the classical 4d K\"ahler potential for K\"ahler moduli $T_i$, $i=1,...,h^{1,1}$, and dilaton $S$ reads~\cite{Grimm:2004uq, Becker:2002nn}
\begin{equation}\label{eq:Kahlerclassic}
K = -2 \ln\hat{ \mathcal V} - \ln{\left( S+ \bar S \right)}\,,
\end{equation}
in units where $M_\text{P} = 1$. Here, $S=e^{-\phi}+iC_0$ denotes the type IIB axio-dilaton, and \linebreak $\hat{\mathcal V} = \int_{X_3} J \wedge J \wedge J$ is the volume of the internal Calabi--Yau threefold $X_3$ with K\"ahler form $J$, which is a function of the two-cycle volumes $v_i$,
\begin{equation}\label{eq:CYvolume}
\hat{\mathcal V} = \frac16 \kappa_{ijk} \hat v_i \hat v_j \hat v_k = \gamma \left(T+\overline T \right)^{\frac32}\,,
\end{equation}
where $\kappa_{ijk}$ is the triple intersection number of the two-cycles whose volume is parameterized by $t_i=v_i+i b_i$, with $\hat v_i=v_ie^{-\phi/2}$. The last equality in eq.~\eqref{eq:CYvolume} holds for the simple case of a single complex K\"ahler modulus $T=t+ i \tau$, which we assume throughout this paper\footnote{In general, the four-cycle volumes, i.e., the real parts of the K\"ahler moduli are defined by $\text Re \, T_i = \frac{\partial \hat{\mathcal V}}{\partial v_i}$.}. In this case $\gamma$ is given by the triple self-intersection $\kappa$ of the two-cycle as $\gamma = \sqrt{3 / 4 \kappa}$. Furthermore, we assume that all complex structure moduli have been stabilized by fluxes~\cite{Giddings:2001yu}.

\medskip

In order to stabilize the remaining K\"ahler modulus we employ quantum corrections to the K\"ahler potential as well as non-perturbative contributions to the superpotential. The latter, in our case, is given by
\begin{equation}\label{eq:generalsuperp}
W = W_0 + \displaystyle\sum\limits_{i=1}^n A_i \, e^{-a_i T}\,,
\end{equation}
where $W_0$ is the vev of the Gukov--Vafa--Witten superpotential~\cite{Gukov:1999ya} determined by fluxes, and the non-perturbative terms are generated by gaugino condensation on stacks of D-branes wrapping the four-cycle parameterized by $T$. The parameters $A_i$ are determined by fluxes and chosen to be real, and the $a_i$ are given by $a_i = \frac{2 \pi}{N_i}$, where $N_i$ is the rank of the $i$-th condensed gauge group.

\medskip

For the K\"ahler potential, we consider two different types of quantum corrections. First, the well-established $\alpha'^3$ corrections derived in~\cite{Grimm:2004uq, Becker:2002nn}. Second, we take recently derived $\alpha'^2$ corrections into account~\cite{Grimm:2013gma}, which will be dubbed GSW corrections. Stemming from a higher-derivative correction to the M-theory action, the $\alpha'^2$ correction for the 4d low-energy effective action of type IIB can be inferred using M-/F-Theory duality and then taking Sen's orientifold limit. The result corresponds to a quantum correction to the volume of $X_3$,
\begin{equation}\label{eq:volumecorr}
\hat{\mathcal V}_q = \hat{\mathcal V} - \frac{5}{64}\mathcal V_{\text{D7} \cap \text{O7}}\,,
\end{equation}
where $\mathcal V_{\text{D7} \cap \text{O7}}= 8 \int_{X_3} c^2_1 (B_3) \wedge J_{B_3}$. Here, $c_1(B_3)$ and $J_{B_3}$ are the first Chern class and the K\"ahler form of the F-theory base manifold $B_3$, respectively.

Thus, after integrating out the dilaton supersymmetrically, the full K\"ahler potential is given by
\begin{equation}\label{eq:Kahlercorr}
K = -2 \ln \left( \hat{\mathcal V}_q + \frac{\hat \xi}{2} \right)\,,
\end{equation}
with $\hat \xi = \xi (S_0 + \bar S_0)^{\frac32}$ in terms of the dilaton vev $S_0$. For our specific case of a single K\"ahler modulus, this translates to
\begin{equation}\label{eq:Kahler1mod}
K = -2 \ln \left[ \sqrt{T+\overline{T}} \left((T+\overline{T})-\frac{15}{8}k^2\right)+ \frac{\hat{\xi}}{2} \right]\,,
\end{equation}
where 
\begin{align}\label{eq:k}
k=\frac{1}{\sqrt 3} \left[ \int_{B_3} c_1 (B_3) \wedge c_1 (B_3) \wedge c_1 (B_3) \right]^{\frac13}\in \mathbbm R\,.
\end{align}
After integrating out $S$, the full F-term scalar potential is determined by eq.~\eqref{eq:generalsuperp} and eq.~\eqref{eq:Kahler1mod} as follows\footnote{Note that since we integrate out $S$, $K^{T \overline T}$ is not the same as the inverse K\"ahler metric of $K=-2 \ln \big(\hat{\mathcal{V}} + \frac{\xi}{2}\big)$ with some constant $\xi$. However, the difference appears only at second order in $\xi/\hat{\mathcal{V}}$.},
\begin{equation}\label{eq:generalscalarp}
V = e^K \left( K^{\overline T T}\overline{D_TW}D_TW-3 |W|^2 \right)\,.
\end{equation}
Note that $V$ is invariant under the following rescaling,
\begin{equation}\label{eq:Vscaling}
\hat{\xi} \longrightarrow \lambda^\frac{3}{2} \hat{\xi}\,, \quad a_i \longrightarrow \frac{a_i}{\lambda}\,, \quad A_i \longrightarrow \lambda^\frac{3}{2} A_i\,, \quad W_0 \longrightarrow \lambda^\frac{3}{2}W_0\,, \quad t \longrightarrow \lambda t\,, \quad k \longrightarrow \lambda^{\frac{1}{2}}k\,,
\end{equation}
for any $\lambda \in \mathbbm R$.  Since the slow-roll predictions discussed below are not altered by this rescaling, we can in principle generate a large number of models at any desired value of $t$. This, however, comes at the potential price of unrealistically large gauge group ranks.

%
%%
%%%
%%%%
%%%%%%%%%%%%%%%%%%%%%%%%%%%%%%%%%%%%%%%%%%%%%%%
%%%%%%%%%%%%%%%%%%%%%%%%%%%%%%%%%%%%%%%%%%%%%%%
\subsection{No-go theorem and large volume expansion}
\label{sec:nogo}
Determining if a model with scalar potential eq.~\eqref{eq:generalscalarp} can accommodate inflation and subsequently produce a de Sitter minimum is a non-trivial task. A classical string tree-level K\"ahler potential of the form $K = -A \ln{ \left(T + \overline T \right)}$ generically  produces an AdS minimum~\cite{Brustein:1992nk}.  A dS minimum can only be achieved by invoking some kind of uplifting mechanism, like inserting anti-D3-branes as in KKLT~\cite{Kachru:2003aw}, or a D-term uplifting mechanism as proposed in~\cite{Parameswaran:2006jh}. More recently, the general case of $0\leq A \leq3$ has been investigated in~\cite{BenDayan:2008dv, Brustein:2004xn, Badziak:2008yg}, resulting in a no-go theorem for both inflation and the existence of a dS minimum in the absence of an uplifting term, for any holomorphic superpotential. This has been formulated as necessary conditions for inflation and dS vacua in~\cite{Covi:2008ea, Covi:2008cn}\footnote{Obviously, the condition and no-go theorems are relevant for supergravity in general, not necessarily string derived/inspired models. For a specific example, see~\cite{BenDayan:2010yz}.}. In particular, the sectional curvature of field space has to fulfill
\begin{equation}\label{eq:nogo1}
R_{i \bar{j}m\bar{n}}f^if^{\bar {j}}f^m f^{\bar {n}}< \frac23\,,\qquad f^i = \frac{G^i}{\sqrt{G^i G_i}}\,,
\end{equation}
in the inflationary region of field space or in the vicinity of the minimum for inflation or a dS minimum to be possible, respectively. Here, $G^i = K^{i \bar j}D_{\bar j} \overline W$ and $G=K+\ln{|W|^2}$. In a regime where both $\alpha'$ corrections are under control, meaning that $\hat \xi \ll \hat{\mathcal V}$ and $k \ll t$, expanding $R$ at large $t$ yields
\begin{equation}\label{eq:nogo2}
R=\frac{2}{3}+\frac{5}{32t^2}\left(5k^4-\frac{7 \hat{\xi}}{3\sqrt{2}\gamma}\sqrt{t}\right).
\end{equation}
Note that this expression is invariant under the rescaling \eqref{eq:Vscaling}.  $k$ is real, so the GSW correction always contributes positively while the $\alpha'^3$ correction can contribute with either sign.
Thus, the new correction makes achieving inflation and/or a dS vacuum even more challenging than before, cf. the discussion in~\cite{Pedro:2013qga}. The necessary condition requires the bracketed expression to be negative, while $k$ appears with fourth power compared to the volume which grows mildly as $\sqrt{t}\sim \mathcal{\hat V}^{1/3}$. Therefore, after including the GSW correction the volume has to be parametrically larger than in models without the correction. 
In the past, suitable dS minima with $\hat{\xi}>0$ and $k=0$ have been constructed~\cite{Rummel:2011cd, deAlwis:2011dp}. In section~\ref{sec:models} we show that both inflation and the existence of dS vacua are possible for certain superpotentials, with $\hat \xi>0$ and even $k \neq 0$ in several benchmark models. 
%
%%
%%%
%%%%
%%%%%%%%%%%%%%%%%%%%%%%%%%%%%%%%%%%%%%%%%%%%%%%
%%%%%%%%%%%%%%%%%%%%%%%%%%%%%%%%%%%%%%%%%%%%%%%
\subsection{CMB observables of interest}

Before discussing our benchmark models, let us review the most important cosmological quantities to be calculated in each model. It is important to note that the full power spectrum is the actual observable, and it is this quantity that should be compared to the data. The usual slow-roll approximations are only simplified parameterizations, useful for a preliminary examination of a model, after which a better statistical analysis is required to fully assess its validity. \\
With a canonically normalized inflaton, we can introduce the usual slow-roll parameters of the scalar potential~\cite{Ade:2013uln},
\begin{equation}\label{eq:slowrollparms}
\epsilon = \frac{1}{2}\frac{V'^2}{V^2}\,, \quad \eta = \frac{V''}{V^2}\,, \quad \xi^2 = \frac{V' V'''}{V^2}\,, \quad \varpi^3 = \frac{V'^2 V''''}{V^3}\,,
\end{equation}
where a prime denotes derivation with respect to the inflaton. Note that the higher order slow-roll parameters $\xi^2$ and $\varpi^3$ can have any sign, and superscripts are not powers but simply denote the order of the slow-roll parameters. We choose this notation to be consistent with the PLANCK collaboration~\cite{Ade:2013uln} and previous literature. The slow-roll approximated power spectrum $P(k)$ is given by~\cite{Ade:2013uln}
\begin{equation}\label{eq:powerspec}
P(k) = \frac{1}{24 \pi^2} \frac{V}{\epsilon}\,.
\end{equation}
Moreover, the scalar spectral index $n_s$, tensor to scalar ratio $r$, running of the spectral index $\alpha$, and the running of the running of the spectral index $\beta$ can be expressed in terms of the above slow-roll parameters as follows~\cite{Ade:2013uln},
\begin{align}\label{eq:cosmobs}
n_s &= 1 + \frac{\text{d} \ln{P(k)}}{\text{d} \ln{k}} \approx 1 + 2 \eta - 6 \epsilon \nonumber \,, \\
r &= 16 \epsilon \nonumber \,,\\
\alpha &= \frac{\text{d} n_s}{\text{d} \ln k} \approx 16 \epsilon \eta - 24 \epsilon^2 - 2 \xi^2 \nonumber\,, \\
\beta &= \frac{\text{d}^2 n_s}{\text{d} \ln k^2} \approx -192 \epsilon^3 + 192 \epsilon^2 \eta - 32 \epsilon \eta^2 - 24 \epsilon \xi^2 + 2 \eta \xi^2 + 2 \varpi^3 \,.
\end{align}
Note that the correct expression for $\alpha$ has an overall minus sign compared to the expression used in~\cite{Ade:2013uln}. All of these CMB observables are evaluated $60$ e-folds before the end of inflation, and we refer to the corresponding field value as the CMB point. More specifically, the CMB point is found by solving the equations of motion \eqref{eq:eom2}, \eqref{eq:eomreal1}, and \eqref{eq:eomreal2} numerically with zero initial velocities.
When running is significant, the above expression for $n_s$ has to be evaluated to the next order in the slow-roll approximation, i.e.,
\begin{equation}\label{eq:nshigherorder}
n_s = 1 + 2\eta - 6\epsilon + 2 \bigg[\frac{1}{3} \eta^2 + (8 C - 1) \epsilon \eta - \bigg(\frac{5}{3} - C \bigg) \epsilon^2 - \bigg( C - \frac{1}{3} \bigg) \xi^2 \bigg]\,,
\end{equation}
with $C = -0.73$ \cite{Lyth:1998xn}.
These expressions are the ones we use to calculate the observables of our benchmark models in section~\ref{sec:models}, and we have verified our results using the full numerical solution of the equations of motion.

CMB and large scale structure (LSS) observations only probe the wave-numbers \linebreak ${H_0\lesssim k\lesssim 1\,\text{Mpc}^{-1}}$ of the power spectrum, which correspond to the first $\sim\! \!8.4$ e-folds from the CMB point. Constraints on smaller scales/larger wave-numbers play a crucial role in further constraining inflationary models. In the early universe, the energy stored in small-scale density perturbations is quickly dissipated by Silk damping, a process that generates $\mu$-type and $y$-type spectral distortions in the CMB black body spectrum. These signals are within observational reach in the near future, for example, by PIXIE~\cite{Kogut:2011xw}. The spectral distortions depend on the shape and amplitude of the primordial power spectrum at wave-numbers ${1\lesssim k \lesssim10^4 \,\text{Mpc}^{-1}}$, which corresponds to e-folds $8.4\,\text{-}\,17.6$ from the CMB point. Therefore, they provide an important model independent constraint on the power spectrum at scales inaccessible by the CMB and LSS observations.

Current constraints on spectral distortions are only upper bounds given by COBE/FIRAS \cite{Fixsen:1996nj}, ARCADE \cite{Seiffert:2009xs} and TRIS \cite{Gervasi:2008eb,Zannoni:2008xx}, with $\mu\lesssim 6 \cdot 10^{-5}$ and $y\lesssim1.5 \cdot 10^{-5}$. PIXIE is expected to have a $5\sigma$ detection of the $\mu$-type ($y$-type) distortion if $\mu= 5\cdot 10^{-8} \, (y=10^{-8})$ \cite{Kogut:2011xw}. Such sensitivity corresponds to a $\lesssim 2\sigma$ evidence in the case of constant $n_s$ with PLANCK error bars \cite{Chluba:2012gq}. For example, simple single-field models with $n_s=0.96,\alpha=\beta=0$ give $\mu\simeq 1.4 \cdot 10^{-8}$ and $y\simeq 3.3 \cdot 10^{-9}$. Enhancement of the power spectrum on the aforementioned scales will increase the spectral distortions signal, making them a useful probe to test running and running of running of $n_s$. 

Using the approximated expressions derived in~\cite{Chluba:2012we}, one can compute both types of spectral distortions as integrals over~$k$,
\begin{align}\label{eq:specdistortion}
\mu &\approx 2.2 \, \int^\infty_{k_{min}} P_{\zeta}(k) \Bigg [ \textrm{exp} \bigg (-\frac{\hat k}{5400} \bigg ) - \textrm{exp} \bigg (- \bigg [ \frac{\hat k}{31.6} \bigg ] ^2 \bigg ) \Bigg ] \text{d}\ln{k} \nonumber\,, \\
y &\approx 0.4 \, \int^\infty_{k_{min}} P_{\zeta}(k) \, \textrm{exp} \Bigg (- \bigg [ \frac{\hat {k}}{31.6} \bigg ] ^2 \Bigg ) \text{d} \ln{k}\,,
\end{align}
with $P_{\zeta} (k) = 2 \pi^2 \frac{P(k)}{k^3}$, $k_\text{min} \approx 1$ Mpc$^{-1}$, and $\hat k = k$ Mpc.

%
%%
%%%
%%%%
%%%%%%%%%%%%%%%%%%%%%%%%%%%%%%%%%%%%%%%%%%%%%%%
%%%%%%%%%%%%%%%%%%%%%%%%%%%%%%%%%%%%%%%%%%%%%%%
%%%%%%%%%%%%%%				 %%%%%%%%%%%%%%%%%%%%%%%
%%%%%%%%%%%%%	%         section 3        %%%%%%%%%%%%%%%%%%%%%%%
%%%%%%%%%%%%%%				  %%%%%%%%%%%%%%%%%%%%%%%
%%%%%%%%%%%%%%%%%%%%%%%%%%%%%%%%%%%%%%%%%%%%%%%
%%%%%%%%%%%%%%%%%%%%%%%%%%%%%%%%%%%%%%%%%%%%%%%
\section{Benchmark models} \label{sec:models}
%
%%
%%%
%%%%
%%%%%%%%%%%%%%%%%%%%%%%%%%%%%%%%%%%%%%%%%%%%%%%
%%%%%%%%%%%%%%%%%%%%%%%%%%%%%%%%%%%%%%%%%%%%%%%
\subsection{Inflation in the original K\"ahler uplifting setup}
\label{sec:1expmodel1}

\begin{figure}[t]
 \centerline{\includegraphics[width=0.8\textwidth]{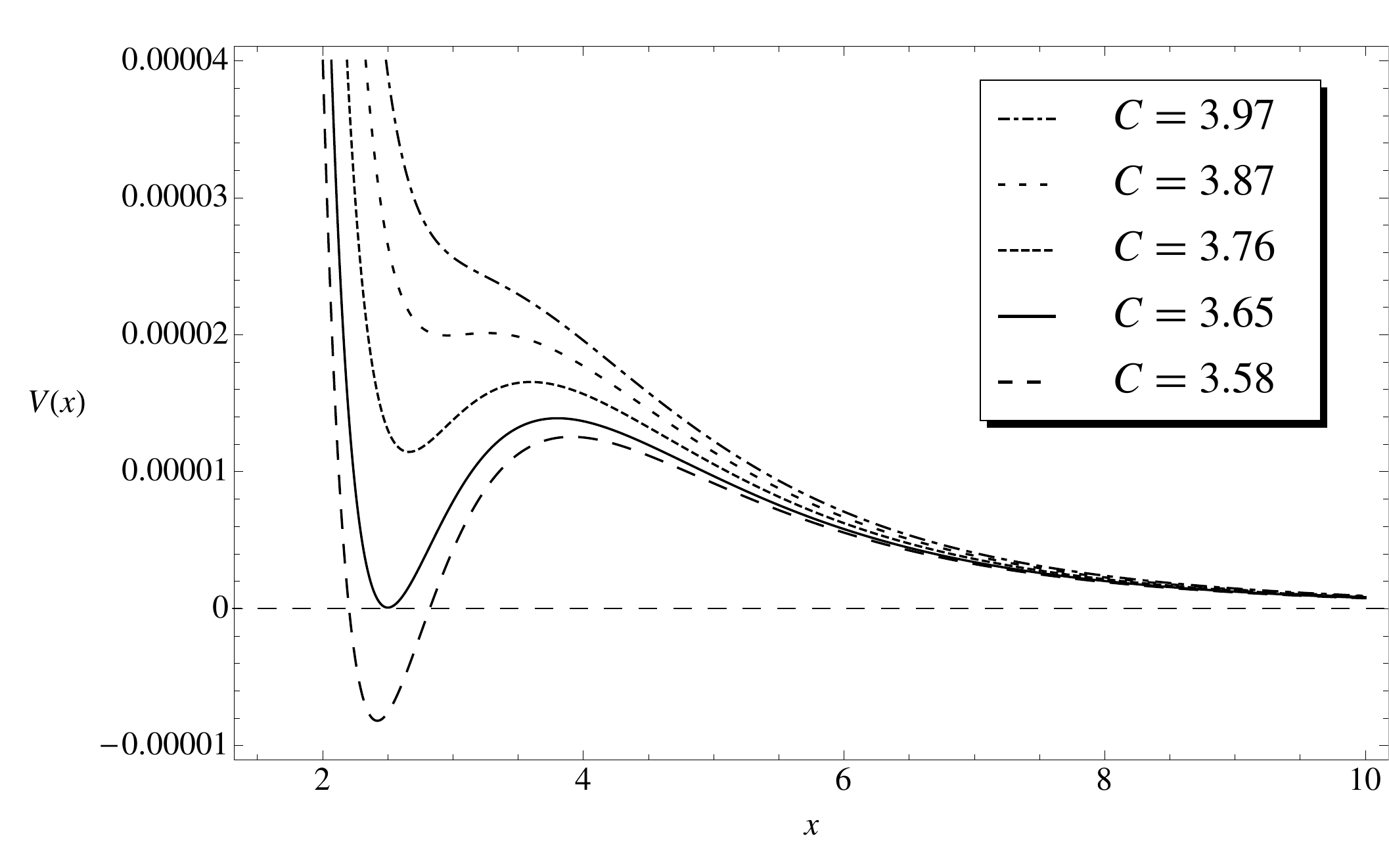}}
 \caption{ The approximated scalar potential along the $x=at$ direction for various values of $C$ (reproduced from~\cite{Rummel:2011cd}).  Evidently, in the vicinity of $C=3.87$ slow-roll inflation from a flat hill-top or inflection point will arise. Fulfilling PLANCK constraints is much more sensitive to $W_0$ than finding generic extrema.}\label{fig:Onemodpot}
\end{figure}

We start our discussion by noting that achieving slow-roll inflation per se is easy in the original setup of K\"ahler uplifting. To see this we consider the original single-modulus example with $k=0$, $\hat\xi>0$ with a single non-perturbative contribution to the superpotential,
\begin{equation}\label{eq:Onemodpot}
V =  e^K \left(K^{T \overline T} \left[ W_T \overline{W_T} + (W_T \cdot \overline{W K_T} + \overline{W_T} \cdot W K_T) \right] + 3 \hat \xi \frac{{\hat \xi}^2 + 7 \hat \xi \hat {\mathcal{V}} + {\hat {\mathcal{V}}}^2}{(\hat {\mathcal{V}} - \hat \xi)(\hat \xi + 2 \hat {\mathcal{V}})^2} |W|^2 \right)\,, 
\end{equation}
where
\begin{equation}
K=-2\ln\left(\hat{\cal V}+\frac{\hat\xi}{2}\right)\,, \qquad W = W_0 + A e^{-a T}\,.
\end{equation}
One can easily show that the K\"ahler modulus axion $\tau$ is minimized at $V_{\tau} = 0$ for $\tau = \frac{n \pi}{a}$ for integer values of $n$.
According to the analysis performed in~\cite{Rummel:2011cd}, in the large volume limit $\hat{\xi}/\hat{\mathcal{V}} \ll 1$ and $A e^{- a t} \ll |W_0|$, the full potential for $t$ and $\tau$ can be considerably simplified to a two-term structure,
\begin{eqnarray}\label{eq:VFt2termttau}
V(t, \tau) \simeq \frac{- W_0 a^3 A}{2 \gamma^2} \left[ \frac{2 C}{9 x^{9/2}} - \frac{e^{-x}}{x^2} \cos(a \tau)\right]\,; \qquad C = \frac{-27 W_0 \hat\xi a^{3/2}}{64 \sqrt{2} \gamma A}\,,
\end{eqnarray}
where $x =  at$. This approximated potential has a meta-stable minimum  at $t \approx 40$ with the following parameters,
\begin{equation}
\label{eqn28}
a = \frac{2 \pi}{100}\,, \quad W_0 = -37.73\,, \quad A = 1\,, \quad \gamma = \frac{\sqrt{3}}{2 \sqrt{5}}\,, \quad \hat{\xi} = 7.98\,.
\end{equation}
Evidently, the value of $C$ determines the vacuum energy at the minimum~\cite{Rummel:2011cd}. The parameter values above correspond to $C\simeq 3.65$, where a dS minimum arises with vacuum energy much smaller than the potential barrier preventing run-away behavior, see figure~\ref{fig:Onemodpot}.
%
%%
%%%
%%%
%%
%
Moreover, from the preceding discussion it is clear that for $\hat{\xi}=0$ the no-go theorem of section~\ref{sec:nogo} applies. Therefore, by changing $C\sim W_0 \hat{\xi}$ continuously, the AdS minimum evolves into a dS minimum and then into a dS inflection point as the threshold to run-away behavior. Again by continuity, we can furthermore fulfill the slow-roll conditions $\epsilon \ll 1$ and $|\eta| \ll 1$ in the vicinity of the value of $C$ generating the dS inflection point. We finally note that adjusting $C$ is done by tuning $W_0$ via the large type IIB flux discretuum. We consider $\hat\xi$ to be fixed as it is defined by the Euler characteristic of a given Calabi--Yau and the desired range of values of the string coupling by flux stabilization of $S$. Figure~\ref{fig:Onemodpot} displays this continuous interpolation from an AdS minimum into an inflationary inflection point graphically.

%
%%
%%%
%%%%
%%%%%%%%%%%%%%%%%%%%%%%%%%%%%%%%%%%%%%%%%%%%%%%
%%%%%%%%%%%%%%%%%%%%%%%%%%%%%%%%%%%%%%%%%%%%%%%
\subsection[Accidental inflation with $k = 0$]{Accidental inflation with $\boldsymbol{k = 0}$}
\label{sec:2expmodel1}

Achieving a stable dS minimum after inflation has ended is not possible using a single non-perturbative term in the superpotential. Therefore, in the remainder of this paper we consider a second gaugino condensate on the same four-cycle, such that 
\begin{equation}\label{eq:Racetracksuperpot}
W = W_0 + A e^{-a T} + B e^{-b T}\,.
\end{equation}
A very simple class of models realizing accidental inflation in string theory is described in~\cite{Linde:2007jn}. There, one uses the above superpotential to generate an AdS KKLT-type SUSY minimum neighboring a second supersymmetric KL-type AdS minimum~\cite{Kallosh:2004yh}. Uplifting this scalar potential with a separate SUSY breaking sector such as an anti-D3-brane led to accidental inflation. 

In the K\"ahler uplifting scenario, a single-field model with $\tau$ stabilized at the origin is obtained for parameter values
\begin{align}\label{eq:model1Rparams}
a = \frac{2 \pi}{12}\,, \quad b = \frac{2 \pi}{41}\,, \quad W_0 = -7.73118337\,, \quad &A = 1\,, \quad B = 0.1598\,, \quad \hat{\xi} = 3.95\,,
\end{align}
and $\gamma = 1$. 
%
%%
%%%
\begin{figure}[t]
\subfigure[]{\includegraphics[width=7cm]{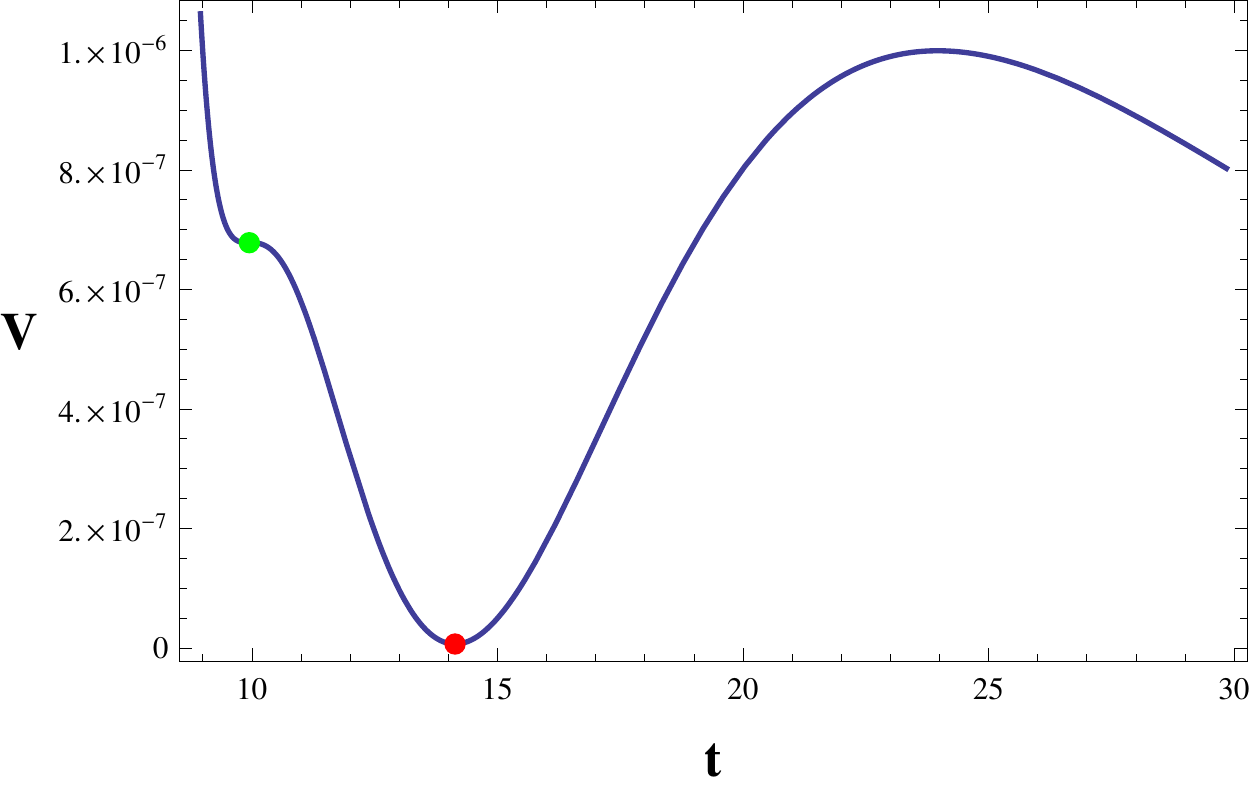}\label{fig:2expmodel1a}}
\subfigure[]{\includegraphics[width=9cm]{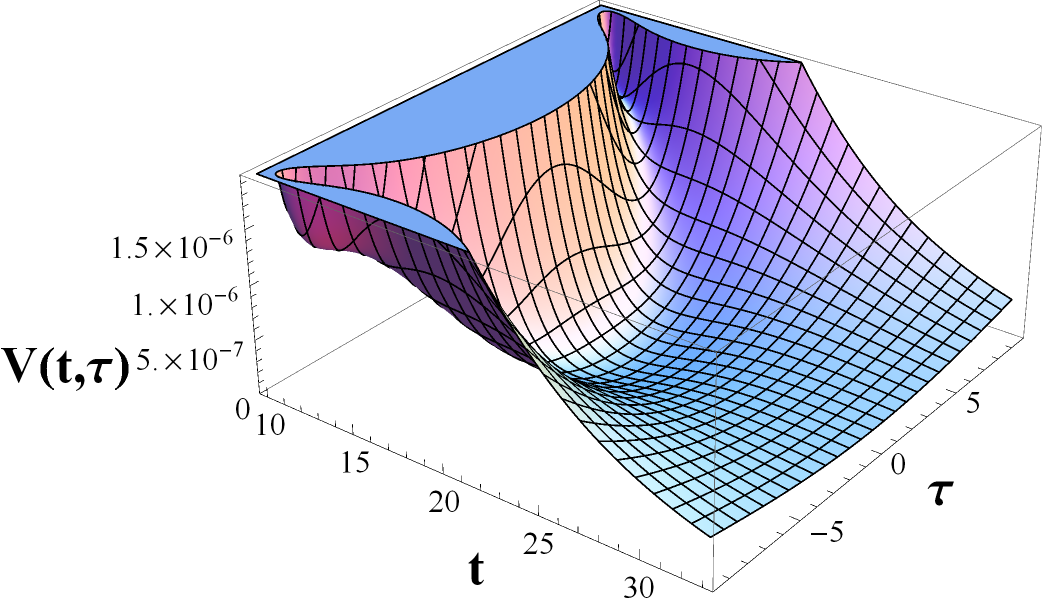}\label{fig:2expmodel1b}}
 \caption{The scalar potential as a function of $t$ for the single-field accidental model with $k = 0$ (a), and for the two-field accidental model (b). In the first example the K\"ahler modulus axion is stabilized at $\tau=0$. The green and red dots mark the locations of the inflection point and the dS minimum, respectively.}\label{fig:2expmodel1}
\end{figure}
%%%
%%
%
As illustrated in figure~\ref{fig:2expmodel1a}, $V$ has an inflection point capable of supporting inflation at $t \approx 10$ and a subsequent dS minimum at $t_\text{min} \approx 14$. Note that the barrier separating the dS minimum from the Dine--Seiberg runaway direction is sufficiently high to guarantee the necessary longevity of the universe. The above parameters were obtained by choosing the CMB point close to $t_\text{CMB} \sim 10$, as well as fixing the gauge group ranks and $\gamma = A = 1$. Then, $W_0$, $B$, $\hat \xi$, and $t_\text{min}$ are inferred by numerically solving the constraints
\begin{align}\label{eq:Vcontraints}
V_T |_{t=t_\text{CMB}} \approx 0\,, \quad \eta |_{t=t_\text{CMB}} \approx -\frac{1}{200}\,, \quad V |_{t=t_\text{min}} \approx  0\,, \quad V_T |_{t=t_\text{min}} \approx 0\,.
\end{align}
Finally, the relevant cosmological observables evaluated at the CMB point are summarized in table~\ref{tab:2expmodels1}.
\begin{table}[t]
\begin{center}
     \begin{tabular}{ccc}
     \toprule
     Parameter & $\quad$ Single-field $\quad$ & $\quad$ Two-field $\quad$ \\
     \midrule
     $n_s$ & 0.974 & 0.963 \\ 
     $r$ & $1.15 \cdot 10^{-11}$ & $3.69 \cdot 10^{-10}$ \\ 
     $\alpha$ & $-2.24 \cdot 10^{-3}$ & $-1.93 \cdot 10^{-3}$ \\ 
     $\beta$ & $-2.91 \cdot 10^{-5}$ & $-3.35 \cdot 10^{-5}$ \\ 
     $\mu$ & $1.74 \cdot 10^{-8}$ & $2.36 \cdot 10^{-8}$ \\ 
     $y$ & $2.44 \cdot 10^{-9}$ & $2.99 \cdot 10^{-9}$ \\ 
     $(t_\text{CMB}, \tau_\text{CMB})$ & $(9.947364,0)$ & $(12.4494, \pm 8.08029)$ \\ 
       \bottomrule
     \end{tabular}
     \caption{Summary of cosmological parameters in the accidental inflation models with $k = 0$.}
     \label{tab:2expmodels1}
\end{center}
\end{table}

\medskip

Using the same functional form of K\"ahler potential and superpotential as in the previous model, but choosing different parameter values gives rise to a very appealing two-field model. Specifically,
\begin{align}\label{eq:model2params}
a = \frac{2 \pi}{30}\,, \!\quad b = \frac{2 \pi}{29}\,, \!\quad W_0 = -1.8041652\,, \!\quad &A = 1\,, \!\quad B = -1.031703\,, \!\quad \hat{\xi} = 1.3\,,
\end{align}
and $\gamma = \sqrt{\frac{3}{20}}$. 
For this parameter choice the inflaton direction is a mixture of the real and imaginary parts of $T$. Two inflection points supporting inflation are found to lie at $t \approx 12.4$ and $\tau \approx \pm 8.1$. The subsequent dS minimum is located on the real axis at $t_\text{min} \approx 20.6$. A three-dimensional plot of $V$ is depicted in figure~\ref{fig:2expmodel1b}. The relevant cosmological observables are again summarized in table~\ref{tab:2expmodels1}. 

%
%%
%%%
%%%%
%%%%%%%%%%%%%%%%%%%%%%%%%%%%%%%%%%%%%%%%%%%%%%%
%%%%%%%%%%%%%%%%%%%%%%%%%%%%%%%%%%%%%%%%%%%%%%%
\subsection{Inflation by deflation}
Remarkably, a slightly different set of parameters produces a single-field inflation model with the novel feature of the inflaton slowly rolling to the left, i.e., the modulus rolls towards smaller volumes during inflation. Henceforth, we call this type of model ``inflation by deflation'' (IBD), since exponential expansion of four-dimensional space-time is achieved by shrinking the volume of the internal manifold.

\begin{figure}[t]
\centering
\includegraphics[width=8cm]{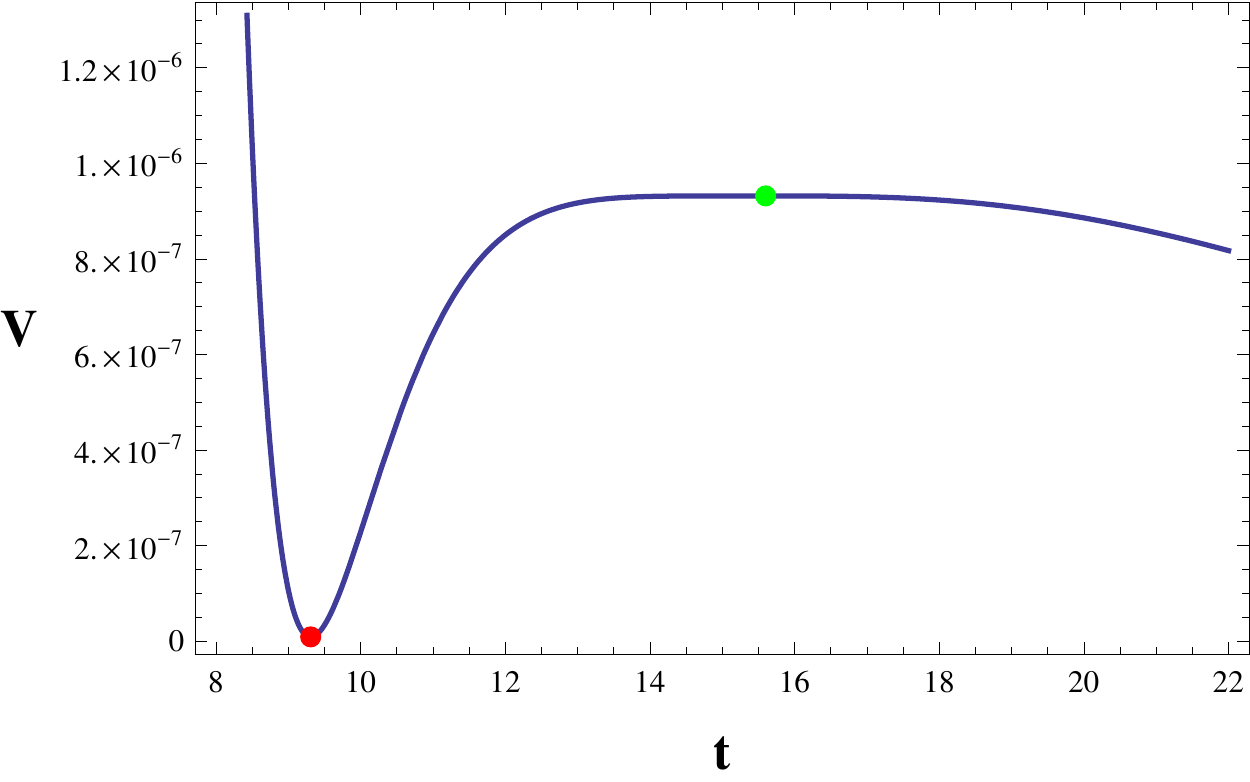}
 \caption{The scalar potential as a function of $t$ for the hilltop IBD model with $k = 0$.  The green and red dots mark the approximate position of the hilltop and $t_\text{min}$, respectively.}
 \label{fig:2exphilltop}
 \end{figure}

This subclass of models avoids two major obstacles of previous string theory setups. First, the overshooting problem simply does not exist since the inflaton rolls towards the infinite barrier at $t \rightarrow 0$. Second, inflation generically starts from a wide barrier between the true dS vacuum and the run-away direction. The barrier is potentially flat enough to guarantee eternal inflation and is insensitive to initial conditions. This allows for model building with a large variety of predictions. In particular, a hilltop IBD model is obtained by choosing
\begin{align}\label{eq:modelhilltopparams}
\quad a = \frac{2 \pi}{12}\,, \quad b = \frac{2 \pi}{37}\,, \quad W_0 = -3.876\,, \quad &A = 1\,, \quad B = 0.161636\,, \quad \hat{\xi} = 6.7\,,
\end{align}
and $\gamma =1$. The predictions of this model are summarized in table~\ref{tab:2expmodelhilltops}. The minimum is located at $t_\text{min} \approx 9.3$, and it is protected by an infinitely high barrier to the left. The corresponding scalar potential as a function of $t$ can be found in figure~\ref{fig:2exphilltop}.

%
%%
%%%
%%%
%%
%

\medskip

\begin{figure}[t]
\subfigure[$V=1 + a_2 \phi^2 - a_n \phi^n$]{\includegraphics[trim=0cm 0cm 3cm 0cm, clip=true, width=8cm]{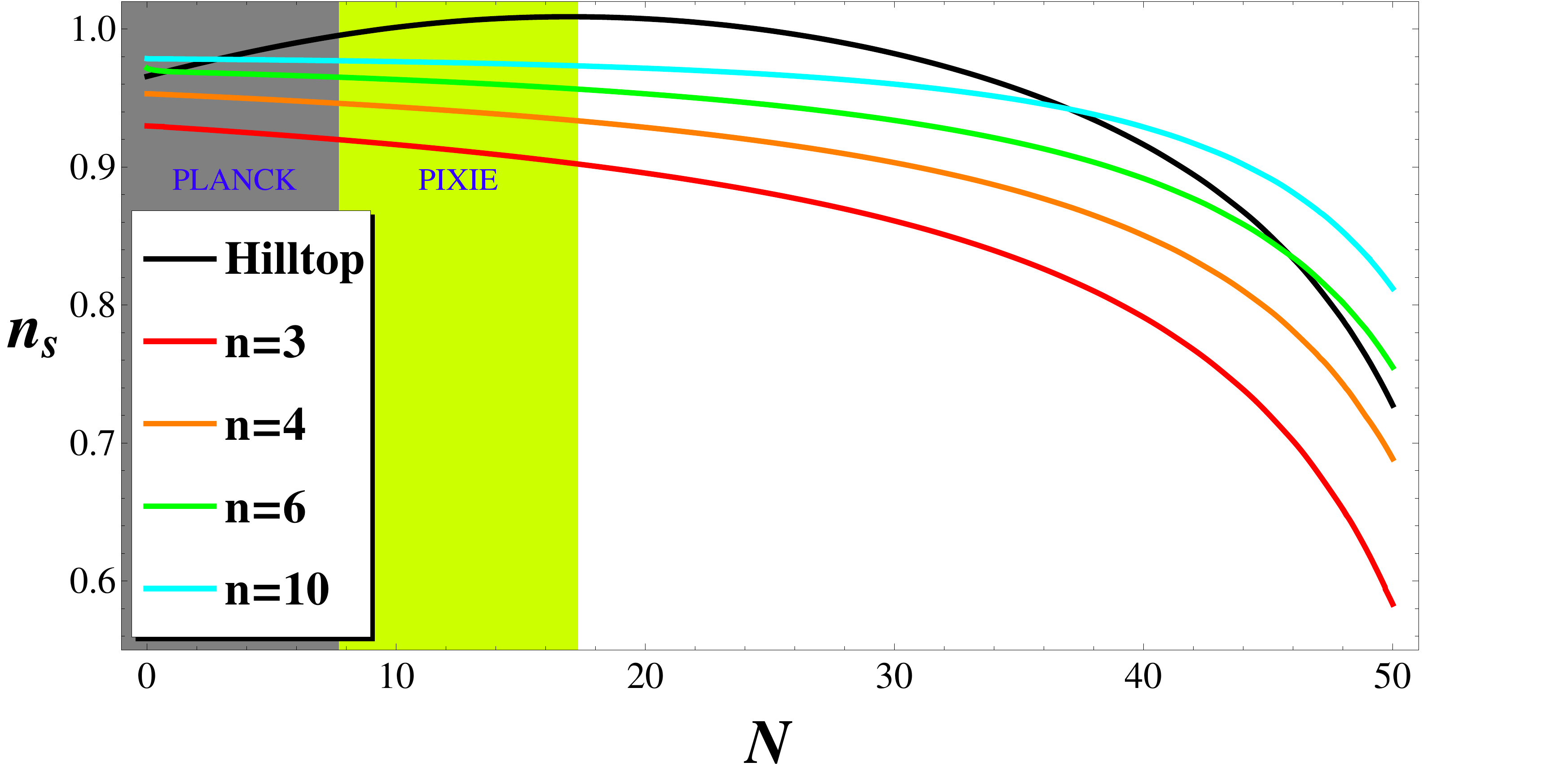}}
\subfigure[$V=1 - a_1 \phi - a_n \phi^n$]{\includegraphics[trim=0cm 0cm 4cm 0cm, clip=true, width=8cm]{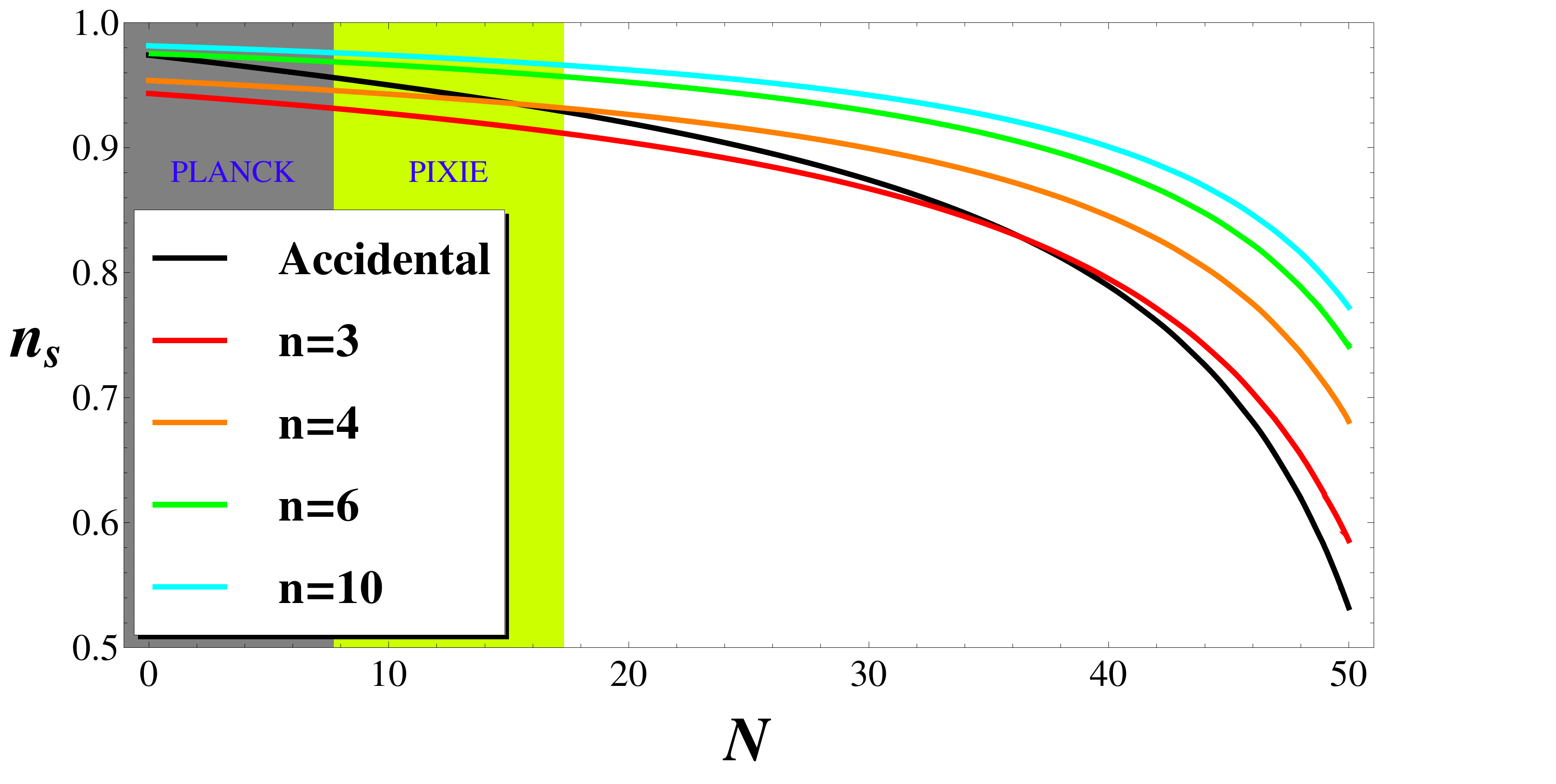}}
\caption{The spectral index $n_s$ as a function of the number of e-folds generated after $t_{CMB}$, with the PLANCK and PIXIE regions of detection indicated. We compare the string-derived IBD model to simple field theory hilltop models (a) $V=1 + a_2 \phi^2 - a_n \phi^n$, and the string-derived accidental inflation model to accidental field theory ones (b) ${V=1 - a_1 \phi - a_n \phi^n}$. } 
\label{fig:2expmodel1nsN}
\end{figure}

Curiously, in this class of models slightly tuning $W_0$ to $-3.817176$, $B$ to $0.161158$, and $\hat \xi = 6.77$ produces a tensor-to-scalar ratio of up to $r \sim 10^{-5}$, which significantly differs from the results of most string inflation models. However, this comes at the price of an increased scale dependence of $n_s$, which forbids raising $r$ even further to an observable level. In a setup with $r \sim 10^{-5}$, $\alpha$ can be as large as 0.012, cf. the summarized observables in table~\ref{tab:2expmodelhilltops}. We remark that an enhanced scale dependence of the spectral index is a distinctive feature of all models under consideration, presumably due to the exponential nature of the employed superpotential. This becomes evident upon plotting $n_s$ as a function of $N$ in comparison to simple field theory models, as illustrated in figure~\ref{fig:2expmodel1nsN}.
\begin{table}[t]
\begin{center}
     \begin{tabular}{ccc}
     \toprule
     Parameter & $\quad$ Hilltop IBD model $\quad$ & Observable running model \\
     \midrule
     $n_s$ & 0.965 & 0.953\\ 
     $r$ & $5.17 \cdot 10^{-6}$ & $1.67 \cdot 10^{-5}$ \\ 
     $\alpha$ & $ 4.72 \cdot 10^{-3}$ & $0.012$\\ 
     $\beta$ & $-1.63 \cdot 10^{-4}$ & $-5.32 \cdot 10^{-4}$\\ 
     $\mu$ & $2.02 \cdot 10^{-8}$ & $2.79 \cdot 10^{-8}$\\ 
     $y$ & $2.45 \cdot 10^{-9}$ & $2.81 \cdot 10^{-9}$\\ 
     $t_\text{CMB}$ & 15.4175 & 15.5058\\ 
     \bottomrule
     \end{tabular}
     \caption{Summary of cosmological parameters in the IBD models with inflation from a hilltop and $k = 0$.}
     \label{tab:2expmodelhilltops}
\end{center}
\end{table}

%
%%
%%%
%%%
%%
%

%
%%
%%%
%%%%
%%%%%%%%%%%%%%%%%%%%%%%%%%%%%%%%%%%%%%%%%%%%%%%
%%%%%%%%%%%%%%%%%%%%%%%%%%%%%%%%%%%%%%%%%%%%%%%
\subsection[Models with $k \neq 0$]{Models with $\boldsymbol{k \neq 0}$}
\label{sec:modelsknon0}

So far we have neglected the GSW correction proportional to $k^2$, cf. eq.~\eqref{eq:Kahler1mod}. However, since $\int_{B_3}c_1 \wedge c_1 \wedge c_1 \neq 0$ in generic type IIB constructions descending from F-theory, considering the case $k \neq 0$ seems unavoidable\footnote{This, however, does not mean that models with $k=0$ are ruled out. As the authors of~\cite{Grimm:2013gma} have remarked, the general structure of perturbative $\alpha'$ corrections to the type IIB effective action is far from understood. It may well be that in the future, corrections are derived which change or cancel the present $\alpha'^2$ correction.}. As illustrated in the following benchmark models with $k=1$, it is still possible to find accidental inflation models and IBD models with a subsequent dS minimum.

%
%%
%%%
\begin{figure}[t!]
\subfigure[]{\includegraphics[width=7.7cm]{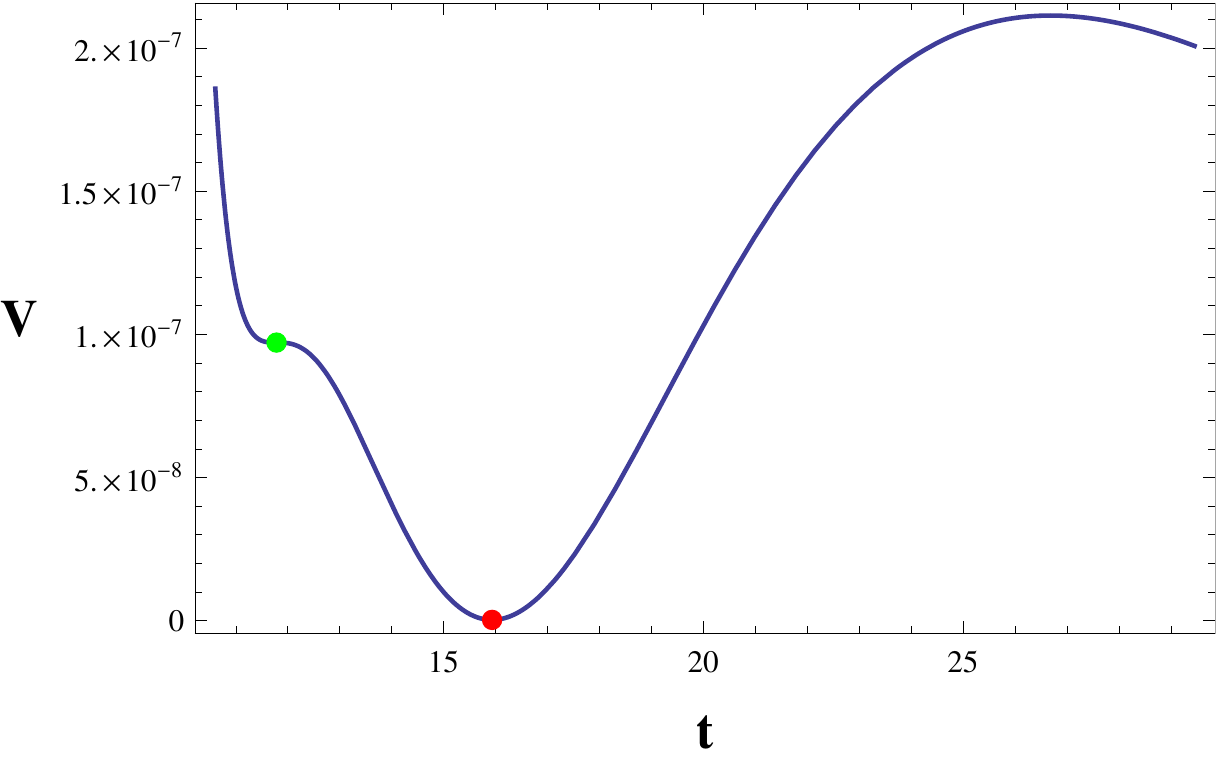}\label{fig:2expmodel3a}}
\subfigure[]{\includegraphics[width=7.7cm]{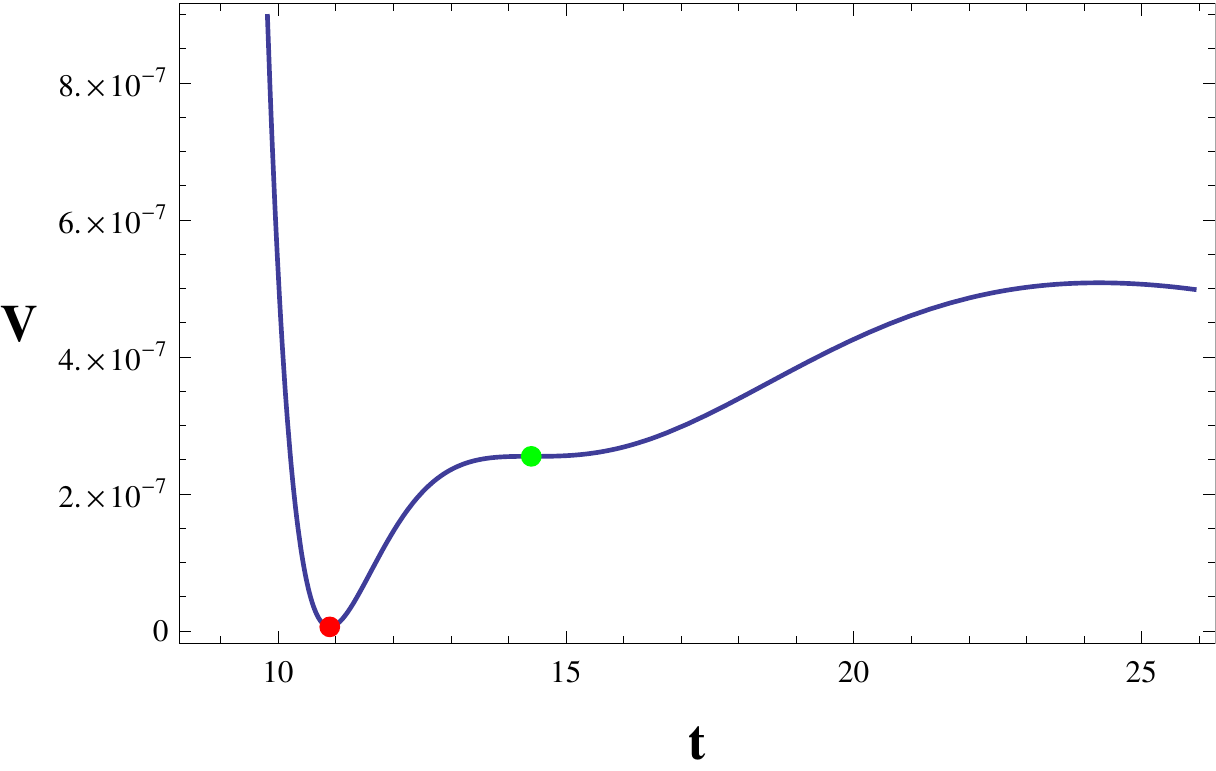}\label{fig:2expmodel3b}}
 \caption{The scalar potential $V$ as a function of $t$ for the first model with $k \neq 0$~(a), and for the second example, which is IBD (b). In both examples the K\"ahler modulus axion is stabilized at $\tau=0$.}\label{fig:2expmodel3}
\end{figure}
%%%
%%
%

In particular, we present two single-field models with sets of closely related parameter spaces. One of them is similar in structure to the first accidental model discussed in section~\ref{sec:2expmodel1}. The other one, as in the $k=0$ case, is an IBD model. Note that for both models, the inclusion of the GSW correction to the K\"ahler potential comes at the price of more fine-tuning for $W_0$ and $B$. We find the following parameters for a model with the inflaton rolling to the right, i.e., accidental inflation from an inflection point,
\begin{align}\label{eq:model3params}
a = \frac{2 \pi}{13}\,, \quad b = \frac{2 \pi}{44}\,, \quad W_0 = -3.242511203\,, \quad B = 0.1163\,, \quad \hat{\xi} =  7\,,
\end{align}
as well as $k= \gamma = A =1$. In this model, inflection point and dS minimum are located on the real axis of $T$ at $t \approx 11.8$ and $t_\text{min} \approx 15.9$, respectively.

\medskip

An IBD-type model is found by slightly changing $b$, $W_0$, $B$, and $\hat \xi$, i.e.,
\begin{align}\label{eq:model3params}
b = \frac{2 \pi}{43}\,, \quad W_0 = -5.1985632\,, \quad B = 0.131344\,, \quad \hat{\xi} =  5.19\,,
\end{align}
with the other parameters unchanged. In this case, inflection point and dS minimum are found on the real axis at $t \approx 14.4$ and $t_\text{min} \approx 10.9$, respectively. The scalar potentials of both benchmark models are displayed in figure~\ref{fig:2expmodel3}. The cosmologically relevant parameters are summarized in table~\ref{tab:2expmodel3params}.

\begin{table}[t!]
\begin{center}
     \begin{tabular}{ccc}
     \toprule
     Parameter & $\quad$ Accidental $\quad$ & $\quad$ IBD $\quad$ \\
     \midrule
     $n_s$ & 0.948 & 0.954 \\ 
     $r$ & $3.48 \cdot 10^{-12}$ & $1.71 \cdot 10^{-10}$ \\ 
     $\alpha$ & $-1.62 \cdot 10^{-3}$ & $-1.71 \cdot 10^{-3}$ \\ 
     $\beta$ & $-4.16 \cdot 10^{-5}$ & $-3.92 \cdot 10^{-5}$ \\ 
     $\mu$ & $1.33 \cdot 10^{-8}$ & $1.35 \cdot 10^{-8}$ \\ 
     $y$ & $1.64 \cdot 10^{-9}$ & $2.05 \cdot 10^{-9}$ \\ 
     $t_\text{CMB}$ & 11.78238504 & 14.3969395 \\ 
     \bottomrule
     \end{tabular}
     \caption{Summary of cosmological parameters in our benchmark models with $k = 1$. In the accidental model the inflaton rolls from left to right, in the IBD model the sides are reversed.}
     \label{tab:2expmodel3params}
\end{center}
\end{table}

%
%%
%%%
%%%%
%%%%%%%%%%%%%%%%%%%%%%%%%%%%%%%%%%%%%%%%%%%%%%%
%%%%%%%%%%%%%%%%%%%%%%%%%%%%%%%%%%%%%%%%%%%%%%%
%%%%%%%%%%%%%%				 %%%%%%%%%%%%%%%%%%%%%%%
%%%%%%%%%%%%%	%         section 4        %%%%%%%%%%%%%%%%%%%%%%%
%%%%%%%%%%%%%%				  %%%%%%%%%%%%%%%%%%%%%%
%%%%%%%%%%%%%%%%%%%%%%%%%%%%%%%%%%%%%%%%%%%%%%%
%%%%%%%%%%%%%%%%%%%%%%%%%%%%%%%%%%%%%%%%%%%%%%%
\section{Conclusions}\label{sec:concl}

We have analyzed models of K\"ahler moduli inflation with a subsequent dS vacuum in the K\"ahler uplifting scenario. In this setup, the generic AdS minimum of the supergravity scalar potential is uplifted to a dS minimum by the interplay of the flux superpotential and the $\alpha'^3$ correction to the K\"ahler potential, parameterized by $\hat \xi$. By means of a geometric necessary condition we have demonstrated that the recently derived GSW correction, which is quadratic in $\alpha'$, is a difficult but surmountable obstacle to model building. On the basis of six benchmark models, we provide examples of different realizations of inflation with a dS minimum, even including the GSW correction.

Specifically, five of our benchmark models have racetrack potentials and either belong to the class of accidental inflation models or ``inflation by deflation'' (IBD) models. Their most important properties can be summarized as follows. All models exhibit a red spectral index $0.94 < n_s < 0.98$ and negligible non-gaussianity, in accord with PLANCK data. The spectral distortions signal is comparable to the standard single-field slow-roll predictions of the simplest models. Furthermore, in all models the gravitino mass is of the same order as the inflationary Hubble scale, and the compactified volume is stabilized at $\hat{\mathcal{V}}_q \sim \mathcal O(100)$. Moreover, the IBD hilltop models stand out regarding their possibly enhanced tensor-to-scalar ratio and enhanced running of the spectral index. If a large enough portion of the power spectrum is probed, this can be used to discern inflation models in the K\"ahler uplifting scenario from simple field theory models. Specifically, the example with $\alpha \sim 0.01$ can probably be tested by the next PLANCK data release. In addition, we find the IBD models to be particularly appealing since they avoid the overshooting problem and ameliorate the initial conditions problem. The width of the flat hilltop, which is $\Delta\phi_T\sim 0.3 M_{\rm P}$ for the canonically normalized inflaton scalar field, raises the question whether some approximate symmetry restoration takes place in this region.

%
%%
%%%
%%%%
%%%%%%%%%%%%%%%%%%%%%%%%%%%%%%%%%%%%%%%%%%%%%%%
%%%%%%%%%%%%%%%%%%%%%%%%%%%%%%%%%%%%%%%%%%%%%%%
\subsection*{Acknowledgements}
We thank Thomas Grimm, Takeshi Kobayashi, Joel Meyers, Francisco Pedro, Fabian R\"uhle, Eric Switzer for interesting discussions. The work of C.W. is supported by a scholarship of the Joachim Herz Stiftung. The work of S.J. is financed by the 2012 undergraduate research grant from the Canadian Institute for Theoretical Astrophysics. The work of I.B.-D. and A.W.  is supported by the Impuls und Vernetzungsfond of the Helmholtz Association of German Research Centres under grant HZ-NG-603, and German Science Foundation (DFG) within the Collaborative Research Center (CRC) 676 �Particles, Strings and the Early Universe�.
%
%%
%%%
%%%%
%%%%%%%%%%%%%%%%%%%%%%%%%%%%%%%%%%%%%%%%%%%%%%%
%%%%%%%%%%%%%%%%%%%%%%%%%%%%%%%%%%%%%%%%%%%%%%%
%%%%%%%%%%%%%%				 %%%%%%%%%%%%%%%%%%%%%%%
%%%%%%%%%%%%%	%         Appendix        %%%%%%%%%%%%%%%%%%%%%%%
%%%%%%%%%%%%%%				  %%%%%%%%%%%%%%%%%%%%%%
%%%%%%%%%%%%%%%%%%%%%%%%%%%%%%%%%%%%%%%%%%%%%%%
%%%%%%%%%%%%%%%%%%%%%%%%%%%%%%%%%%%%%%%%%%%%%%%

\begin{appendix}

%
%%
%%%
%%%%
%%%%%%%%%%%%%%%%%%%%%%%%%%%%%%%%%%%%%%%%%%%%%%%
%%%%%%%%%%%%%%%%%%%%%%%%%%%%%%%%%%%%%%%%%%%%%%%
\section{Generalized slow-roll equations of motion}
\label{sec:Appendixeom}

The generalized supergravity slow-roll parameters are defined by~\cite{Lyth:1998xn,BlancoPillado:2006he,BenDayan:2008dv}:
\begin{subequations}
\begin{align}\label{eqn15}
\epsilon &= \frac{K^{i \bar{j}} \partial_i V \partial_{\bar{j}} V}{V^2}\,,\\
\eta &= min\bigg\{\text{EV}
	\begin{pmatrix}
 	  K^{i \bar{m}} N_{\bar{m} j} & K^{i \bar{m}} N_{\bar{m} \bar{j}} \\
  	K^{\bar{i} m} N_{m j} & K^{\bar{i} m} N_{m \bar{j}}
 	 \end{pmatrix} \bigg\}\,,
\end{align}
\end{subequations}
where
\begin{equation}
\label{eqn17}
N_{i \bar{j}} = \frac{\partial_i \partial_{\bar{j}} V}{V}\,, \quad
N_{i j} = \frac{\partial_i \partial_{j} V - \Gamma_{i j}^l \partial_l V}{V}\,.
\end{equation}
The Christoffel symbol for the K\"ahler moduli space is given by
\begin{equation}
\label{eqn19}
\Gamma_{i j}^l = K^{l \bar{n}} \partial_j \partial_i \partial_{\bar{n}} K\,.
\end{equation}
Inflation ends when $\epsilon=1$. Since we consider a single K\"ahler modulus, $\eta$ is a $2 \times 2$ matrix. \\
The relevant equations of motion for the complex modulus $T$ are given by
\begin{subequations}
\begin{align}\label{eq:eom1}
3H^2 &= K_{T \overline T}|\dot{T}|^2+V\,,\\
\label{eq:eom2}
\dot{H} &= - K_{T\overline T} (\dot t^2 + \dot \tau^2)\,,\\
\label{eq:eom3}
0 & =\ddot{T}+3H\dot{T}+\Gamma^T_{TT}\dot{T}^2+K^{T \overline T}\partial_{\overline T}V \,.
\end{align}
\end{subequations}
In addition, there is the complex conjugate of eq.~\eqref{eq:eom3}. This translates into the equations of motion for $t$ and $\tau$ as follows,
\begin{subequations}
\begin{align}\label{eq:eomreal1}
\ddot{t}+3H\dot{t}+\Gamma^T_{TT}(t)(\dot{t}^2-\dot{\tau}^2)+\frac{K^{T \overline T}(t)}{2}\partial_tV=0 \,, \\
\label{eq:eomreal2}
\ddot{\tau}+3H\dot{\tau}+2\Gamma^T_{TT}(t)\dot{t}\dot{\tau}+\frac{K^{T \overline T}(t)}{2}\partial_{\tau}V=0 \,.
\end{align}
\end{subequations}
Since the K\"ahler modulus is not canonically normalized, the standard expression for the number of e-folds of exponential expansion changes. Assuming that $\tau$ is generically very heavy, the resulting model is effectively a single-field model. In this case eq.~\eqref{eq:eomreal2} simplifies to
\begin{equation}\label{eq:eomreal2simp}
3 H \dot t \simeq - \frac{K^{T \overline T}}{2} \partial_t V(t, \tau)\,.
\end{equation}
Using that $K_{T \overline T} = \frac{1}{4} K_{tt}$ and $\partial _T V \partial _{\overline T} V = \frac{1}{4} \frac{\partial V}{\partial t}^2$, we obtain the following expression for the number of e-folds $N$,
\begin{equation}
\label{eq:nrefolds}
N \approx \frac{1}{2} \int_{t_\text{end}}^{t} \! \, \frac{\sqrt{K_{t t}}}{\sqrt{\epsilon}} \, \mathrm{d} t\,,
\end{equation}
where we have used that $H^2 \approx \frac13 V$. This expression can be verified numerically by solving the exact background dynamics. The additional factor of $\sqrt{K_{t t}}$ is crucial in string theory models because usually $K_{tt}\ll1$, especially in large volume scenarios. Therefore, it is more difficult to generate enough e-folds on the one hand, and enhancing the scalar-to-tensor ratio $r$ on the other hand. This explains why string inflation models usually have a gravitational wave signal even smaller than simple small-field models~\cite{BenDayan:2009kv}.

%
%%
%%%
\begin{figure}[bt]
 \centerline{\includegraphics[scale=1]{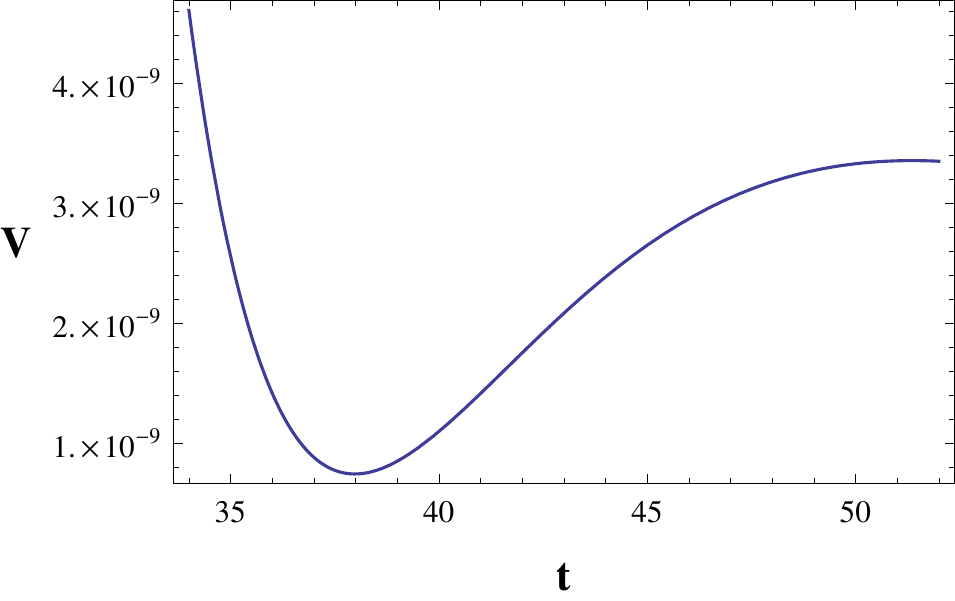}}
 \caption{An example of a dS minimum at large volume with sufficiently high barrier and the new $\alpha'^2$ correction taken into account.}\label{fig:Sminimum}
\end{figure}
%%%
%%
%

%
%%
%%%
%%%%
%%%%%%%%%%%%%%%%%%%%%%%%%%%%%%%%%%%%%%%%%%%%%%%
%%%%%%%%%%%%%%%%%%%%%%%%%%%%%%%%%%%%%%%%%%%%%%%
\section{Meta-stable dS minima at large volume}
\label{sec:Appendixmin}
Realizing a dS minimum at larger values of $t$ without employing unrealistically large gauge groups can be achieved by choosing $a$ and $b$ very close together, as carried out in~\cite{Sumitomo:2013vla}. There, the authors found a dS minimum at $t_\text{min} \approx 42.5$ using $a=\frac{2 \pi}{14}$ and $b=\frac{2 \pi}{15}$, with $k=0$ and $\hat \xi = 3.41 \cdot 10^{-3}$. However, there are two potential issues with this particular solution. First, for such low values of $\hat \xi$ the no-go theorem eq.~$\eqref{eq:nogo2}$ puts a highly unreasonable bound on $k$, i.e. $k \lesssim 0.3$. Second, the barrier separating the dS minimum from the run-away direction is too small to support the necessary longevity of the resulting universe.

An example evading both of these caveats has the following parameter values
\begin{align}\label{eq:Sminimumparams}
a = \frac{2 \pi}{45}\,, \quad b = \frac{2 \pi}{41}\,, \quad W_0 = -0.822\,, \quad &A = 1\,, \quad B = -1.263\,, \quad \hat{\xi} = 16.92\,,
\end{align}
and $k=1$. In this case, the model has a strongly stabilized minimum $V_\text{min} = 7.47 \cdot 10^{-10}$ at $t_\text{min} \approx 38$, corresponding to a volume $\hat{\mathcal{V}}_q \approx 646$. The gravitino mass in this minimum is $m_{3/2}^2 = 1.57 \cdot 10^{-6}$ in Planck units. The corresponding scalar potential is illustrated in figure~\ref{fig:Sminimum}.
\end{appendix}

%
%%
%%%
%%%%
%%%%%%%%%%%%%%%%%%%%%%%%%%%%%%%%%%%%%%%%%%%%%%%
%%%%%%%%%%%%%%%%%%%%%%%%%%%%%%%%%%%%%%%%%%%%%%%
%%%%%%%%%%%%%%				 %%%%%%%%%%%%%%%%%%%%%%%
%%%%%%%%%%%%%	%        Bibliography    %%%%%%%%%%%%%%%%%%%%%%%
%%%%%%%%%%%%%%				  %%%%%%%%%%%%%%%%%%%%%%
%%%%%%%%%%%%%%%%%%%%%%%%%%%%%%%%%%%%%%%%%%%%%%%
%%%%%%%%%%%%%%%%%%%%%%%%%%%%%%%%%%%%%%%%%%%%%%%

\bibliography{Draft_40}
\bibliographystyle{JHEP}

\end{document}